\documentclass[12pt, journal]{IEEEtran}
\usepackage{graphicx}
\usepackage{comment}

\ifCLASSINFOpdf
 
\else
 
\fi

\usepackage{tikz}
\usetikzlibrary{shapes.geometric, arrows.meta, positioning}

\tikzstyle{block} = [rectangle, rounded corners, minimum width=3.5cm, minimum height=1.2cm,text centered, draw=black, fill=blue!10]
\tikzstyle{arrow} = [thick,->,>=stealth]
\usetikzlibrary{fit}

\newcommand{\reff}[1]{Fig.~\ref{#1}}
\newcommand{\reft}[1]{Table \ref{#1}}
\begin{document}

\title{Enabling Continuous 5G Connectivity in Aircraft through Low Earth Orbit Satellites}
%\title{Continuous 5G Connectivity in Aircraft via LEO Satellites: A Simulation-Based Approach}
%
%% para modificar %%
\author{Raúl Parada, Victor Monzon Baeza,~\IEEEmembership{Senior Member,~IEEE}, Carlos Horcajo Fernández de Gamboa, Rocío Serrano Camacho, and Carlos Monzo,~\IEEEmembership{Senior Member,~IEEE}
\thanks{Manuscript received XXX, XX, 2025; revised XXX, XX, 2025.}
\thanks{R. Parada is with Centre Tecnològic de Telecomunicacions de Catalunya (CTTC), Castelldefels, Spain. V. M. Baeza, C. H. F. de Gamboa, R. S. Camacho, and C. Monzo, are with Universitat Oberta de Catalunya (UOC), Barcelona, Spain.  \textit{Corresponding author: Raúl Parada (rparada@cttc.es).}
}}

\markboth{Journal of \LaTeX\ Class Files,~Vol.~XX, No.~XX, April~2025}%
{Shell \MakeLowercase{\textit{et al.}}: Bare Demo of IEEEtran.cls for IEEE Journals}



\maketitle

\begin{abstract}
%The increasing demand for high-speed connectivity in commercial aviation requires innovative solutions to ensure uninterrupted service throughout flights. This paper proposes a flexible and scalable system that utilizes low-earth orbit satellites to provide continuous 5G connectivity to aircraft during flight. The system is designed to accommodate additional satellite constellations and ground stations, ensuring redundancy and scalability. In addition, a dedicated 5G transmission and reception (Tx/Rx) distribution network within the aircraft improves signal quality for passengers. The simulation results validate the feasibility of this approach, demonstrating extended coverage along flight routes and improved in-cabin connectivity.

As air travel demand increases, uninterrupted high-speed internet access becomes essential. However, current satellite-based systems face latency and connectivity challenges. While prior research has focused on terrestrial 5G and geostationary satellites, there is a gap in optimizing Low Earth Orbit (LEO)-based 5G systems for aircraft. This study evaluates the feasibility of deployment strategies and improving signal quality with LEO satellites for seamless in-flight 5G connectivity. Using Matlab and Simulink, we model satellite trajectories, aircraft movement, and handover mechanisms, complemented by ray-tracing techniques for in-cabin signal analysis. Results show that proposed LEO satellite configurations enhance coverage and reduce latency, with sequential handovers minimizing service interruptions. These findings contribute to advancing in-flight 5G networks, improving passenger experience, and supporting real-time global connectivity solutions.

\end{abstract}


\IEEEpeerreviewmaketitle

\section{Introduction}\label{intro}
\subsection{Motivation}
New communication systems generations, such as 5G, introduce new functionalities for aircraft operations. This results in a growing demand for in-flight connectivity for entertainment, work, and safety, which has exposed the limitations of traditional aircraft communication systems, particularly those relying on geostationary satellites (GEO) \cite{darwish2022leo}. These systems struggle with high latency and limited coverage, especially over remote and oceanic regions. The emergence of non-terrestrial networks (NTN), particularly Low Earth Orbit (LEO) satellites, offers a transformative solution \cite{5G_NTN}, and the integration of 5G-NTN is imminent from standardization groups such as 3GPP. LEO satellites, operating at 500-2,000 kilometers, offer lower latency and higher data throughput, enabling real-time applications like video conferencing and online gaming. Their dense deployment boosts network capacity, ensuring seamless global coverage, even across polar and transoceanic routes. Moreover, implementing massive inter-satellite links (ISL) within LEO constellations enhances network resilience and reduces reliance on ground stations \cite{mISL}. ISLs enable direct satellite communication, optimizing data routing and reducing latency. These capabilities are particularly beneficial for in-flight connectivity, where consistent and uninterrupted service is crucial.

Recent advancements underscore the potential of LEO satellites in revolutionizing in-flight connectivity. For instance, companies like OneWeb and SpaceX's Starlink have launched extensive LEO constellations to provide global broadband services, including to aviation sectors. OneWeb, with 600 satellites, provides aviation customers with low-latency (under 50 ms) and high-speed (100+ Mbps) internet. Starlink, with over 5,000 satellites, offers up to 350 Mbps per aircraft and ultra-low latency (around 20 ms), enabling seamless browsing, streaming, and real-time applications for all passengers. These advancements shift how airlines deliver high-speed, reliable onboard internet. 

The transition to LEO-based continuous connectivity in aircraft presents a financial shift where long-term operational savings offset short-term capital investment. Initially, airlines faced increased CAPEX due to the need for upgraded onboard hardware, such as LEO-compatible antennas and networking systems. However, early adopters gain a competitive edge by offering superior passenger experiences and unlocking new revenue streams through premium in-flight connectivity services. In addition, advancements in chipset technology have enhanced onboard satellite processing \cite{ortiz2023onboard} to faster data handling, reduced latency, and optimized bandwidth, ensuring reliable high-speed internet for passengers and crew. Over time, OPEX reductions become more evident as lower data transmission costs, fuel savings from optimized flight routes, and predictive maintenance efficiencies drive down operational expenses. As LEO infrastructure scales and service costs decline, airlines will see a more balanced financial model where high-speed, low-latency connectivity becomes a cost-effective standard rather than a premium add-on. This strategic investment positions airlines for sustained profitability, improved efficiency, and enhanced customer satisfaction in the evolving landscape of connected aviation.

\vspace{-2mm}
\subsection{Contributions}
%The objectives of this paper are:
To ensure continuous 5G connectivity in aircraft via LEO constellations, our key contributions are:
\begin{itemize}
    \item Evaluating the feasibility of using LEO constellations for continuous in-flight 5G coverage, including satellite deployment strategies and efficient handover mechanisms to minimize latency.
    \item Introducing a Tx/Rx distribution system to improve passenger connectivity and network efficiency to assess 5G signal propagation inside aircraft.
    \item Providing recommendations for future advancements in satellite-based 5G for aviation to maintain seamless and continuous connectivity.
\end{itemize}
We envision a continuous 5G connectivity in-flight where passengers can enjoy seamless high-speed services to increase satisfaction. 
    %\item To analyze the feasibility of using LEO constellations for in-flight 5G connectivity.
    %\item To design and simulate satellite deployment strategies that maximize coverage throughout a flight.
    %\item To evaluate the impact of satellite handovers on network performance and latency.
    %\item To assess indoor 5G signal propagation within aircraft cabins and propose optimizations.
    %\item To provide recommendations for future enhancements in satellite-based 5G networks for aviation.
    %\item A system enabling uninterrupted 5G connectivity via LEO satellites during cruise phases.
    %\item A flexible architecture allowing integration of additional satellites and terrestrial stations.
    %\item A novel in-cabin 5G Tx/Rx distribution system to enhance the passenger experience.
    
This paper is organized as follows. Section \ref{sec:bg} provides an overview of related work in the field. Section \ref{sec:meth} details the simulation environment, satellite configurations, and in-cabin signal analysis. Section \ref{sec:exp} presents findings on coverage improvement, in-cabin signal quality, and performance comparisons.
Finally, Section \ref{sec:conc} summarizes key insights and suggests directions for future research.

\vspace{-1mm} 
%\hfill January 6, 2025
\section{5G-LEO for in-flight Connectivity} \label{sec:bg}
Recent literature extensively explores the integration of 5G systems in aviation, mainly through LEO satellite networks. A 5G network architecture leveraging LEO satellites for backhaul connectivity in passenger flights is examined in \cite{mafakheri2023edge}, highlighting the role of Multi-access Edge Computing (MEC) in optimizing in-flight services through machine learning-based enhancements. Similarly, \cite{albagory2020modelling} compares a 5G-based aircraft-stratospheric base station network with LEO satellites Internet, demonstrating key advantages such as lower latency, reduced atmospheric attenuation, extended link duration, and minimal handover interruptions, factors that significantly enhance aviation communication reliability.

Beyond LEO satellites, NTN extends beyond traditional aircraft communication with the use of unmanned aerial vehicles (UAVs) in conjunction with LEO to explore 5G Internet of Remote Things (IoRT) applications \cite{ma2021uav}. Their study focuses on efficient data collection and resource allocation in infrastructure-free environments, demonstrating how LEO connectivity supports remote and dynamic networking scenarios. Furthermore, \cite{leon2024towards} analyzes the incorporation of 5G NTN communications in manned and unmanned rotary-wing aircraft. They address challenges such as low-altitude flight operations and blade interference while emphasizing the advantages of multi-orbital connectivity. In addition, 5G New Radio (NR) communication over LEO constellations to support connectivity for mobile platforms such as aircraft is investigated in \cite{maattanen20195g}. This study underscores the role of LEO networks in expanding 5G coverage in remote regions while improving overall communication capabilities.

%These studies highlight that LEO-based 5G architectures present a viable solution for delivering high-performance in-flight communication services and bridging connectivity gaps in remote airspace by addressing critical aspects such as latency reduction, seamless handovers, and enhanced connectivity.

Despite significant advancements in integrating 5G-LEO satellite networks for aviation, existing studies focus on specific aspects such as network architecture, UAV-LEO integration, multi-orbital connectivity, and general broadband applications. However, a critical gap remains in evaluating continuous 5G connectivity for aircraft throughout an entire flight, particularly in handover management, seamless coverage optimization, and in-cabin network performance. While previous research has demonstrated the feasibility of LEO-based NTN for aviation, few studies provide a holistic approach that models, simulates, and validates real-time in-flight connectivity continuity under dynamic flight conditions. This work addresses these shortcomings by presenting an LEO-based 5G architecture to ensure uninterrupted connectivity through improved satellite handover mechanisms, adaptive deployment strategies, and enhanced in-cabin 5G signal distribution. This paper's findings will provide a more robust and scalable framework for next-generation in-flight broadband services. Overall, we present a simulation framework based on Matlab and Simulink tools to provide a continuous 5G in-fight connectivity along an LEO constellation.

\section{System Architecture} \label{sec:meth}

The proposed system architecture aims for a modular software framework to deliver continuous 5G connectivity to aircraft in flight using a layered approach that integrates both LEO satellite constellations and ground stations (GS) with an intelligent in-cabin signal distribution network. Key functionalities include (i) maintaining seamless handover between satellites during flight, (ii) maximizing signal reliability within the aircraft cabin, and (iii) supporting future scalability via modular integration of satellites and GS. This architecture integrates dynamic satellite management based on flight paths with ray-tracing-based antenna placement to ensure passengers experience uninterrupted, high-quality service throughout the journey.

%Ensuring seamless 5G connectivity in aircraft requires a sophisticated and multi-layered approach that integrates both space-based and terrestrial technologies. The architecture must address challenges such as dynamic handover between satellites, latency management, and optimizing signal distribution within the aircraft. The proposed system leverages LEO satellite networks to provide continuous connectivity, ensuring high-speed and low-latency communication throughout the flight duration. Furthermore, the system incorporates mechanisms for scalability, allowing the addition of new satellite constellations and ground stations as network demand evolves. Lastly, the on-board distribution system ensures that passengers and crew members receive a stable and high-quality 5G connection, regardless of their location within the aircraft. Figure \ref{fig:scheme} describes how the Simulink workflow works by indicating the interaction between elements. This workflow is an improvement by the official scheme created by Mathworks \cite{mathworks_spacecraft_attitude}.
\vspace{-2mm}
\subsection{Simulink-Based Orbital Communication Framework}

\begin{figure}[!t]
\centering
\includegraphics[width=1\linewidth]{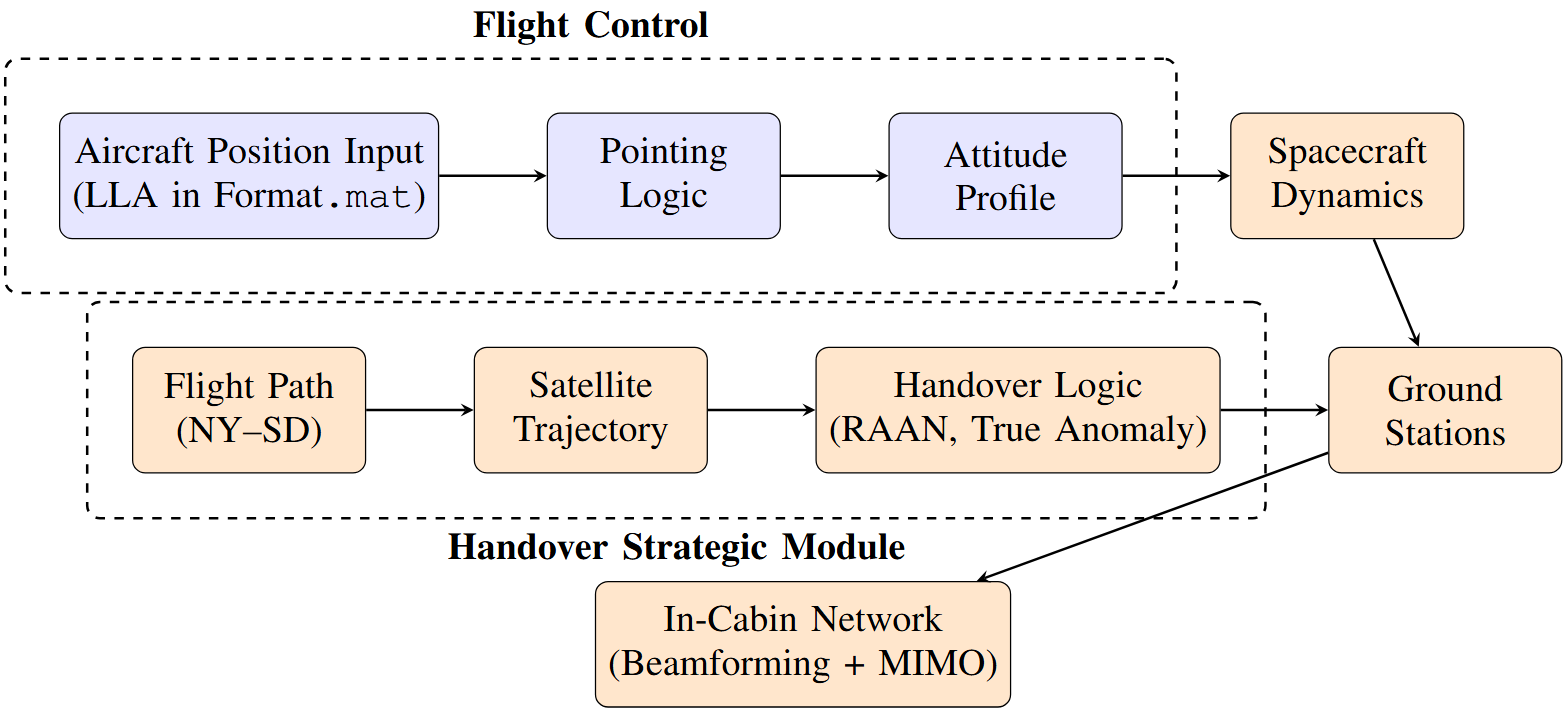}
\caption{System architecture for continuous 5G connectivity in aircraft using LEO satellites.}
\label{fig:system_architecture}
\end{figure}

A modular framework was implemented using MATLAB and Simulink to simulate satellite-based connectivity. \reff{fig:system_architecture} illustrates the overall system architecture designed to enable continuous 5G connectivity for aircraft using LEO satellites. The diagram is inspired by the official simulation flow proposed by MathWorks for spacecraft attitude analysis \cite{mathworks_spacecraft_attitude}, and includes inherited components and new contributions highlighted in orange. New or modified blocks are described in \reft{TableSystem}. 
\begin{table}[h]
\caption{Parameter Blocks Description for Simulation Workflow}
\centering
\begin{tabular}{|l|p{6cm}|}
\hline
\multicolumn{1}{|p{1.7cm}|}{\textbf{Module}} & \multicolumn{1}{|p{6.3cm}|}{\textbf{Description}} \\ \hline
\multicolumn{1}{|p{1.7cm}|}{Aircraft Position Input} & \multicolumn{1}{|p{6.3cm}|}{Inputs the aircraft's dynamic position (latitude, longitude, altitude (LLA)) in Format\texttt{.mat} file.} \\ \hline
\multicolumn{1}{|p{1.7cm}|}{Pointing Logic} & \multicolumn{1}{|p{6.3cm}|}{Computes the satellite orientation relative to the aircraft’s position.} \\ \hline
\multicolumn{1}{|p{1.7cm}|}{Attitude Profile} & \multicolumn{1}{|p{6.3cm}|}{Translates orientation into attitude commands (pitch, roll, yaw).} \\ \hline
\multicolumn{1}{|p{1.7cm}|}{Spacecraft Dynamics} & \multicolumn{1}{|p{6.3cm}|}{Simulates orbital motion: position, velocity, orientation.} \\ \hline
\multicolumn{1}{|p{1.7cm}|}{Handover Logic} & \multicolumn{1}{|p{6.3cm}|}{Manages dynamic LEO handovers to maintain uninterrupted 5G coverage during flight.} \\ \hline
\multicolumn{1}{|p{1.7cm}|}{In-Cabin Networks} & \multicolumn{1}{|p{6.3cm}|}{Models and analyzes in-cabin 5G with ray-tracing and MIMO to evaluate signal quality.} \\ \hline
\end{tabular}
\label{TableSystem}
\end{table}

\vspace{-5mm} 
One of the key modifications appears in the \textit{Satellite Trajectory} node, used initially to define observational missions from orbit. In our model, this block has been adapted to represent the trajectory of a LEO satellite that must provide temporary coverage to a moving aircraft and establish a communication link with a terrestrial GS. To achieve this, new Simulink modules were introduced to support satellite state propagation and orientation alignment. Specifically, the \textit{Spacecraft Dynamics} block models the satellite’s position, velocity, and attitude, while the \textit{Attitude Profile} block governs its orientation with respect to detected targets (e.g., aircraft or GS). Together, these elements enable a more realistic and functional simulation of satellite-environment interactions, addressing limitations in the original reference framework.
The \textit{Ground Stations} node enables the inclusion and distribution of multiple GS under different strategic configurations, as detailed in \cite{Victor} to support flexible terrestrial connectivity scenarios essential for robust satellite-aided in-flight service.

Finally, two additional modules—\textit{Handover Logic} (III-B) and \textit{In-Cabin Network} (III.C)—were introduced to support continuous 5G service. These two blocks are essential to the core contribution of this work and are discussed in detail in the following sections.

%These orange-colored blocks represent novel additions to the standard workflow that enhance satellite modeling by flight dynamics and orbital propagation models to support real-time tracking and connectivity analysis. 

%\input{Diagram/Flowchart}
\begin{figure}[!t]
\centering
\includegraphics[width=1\linewidth]{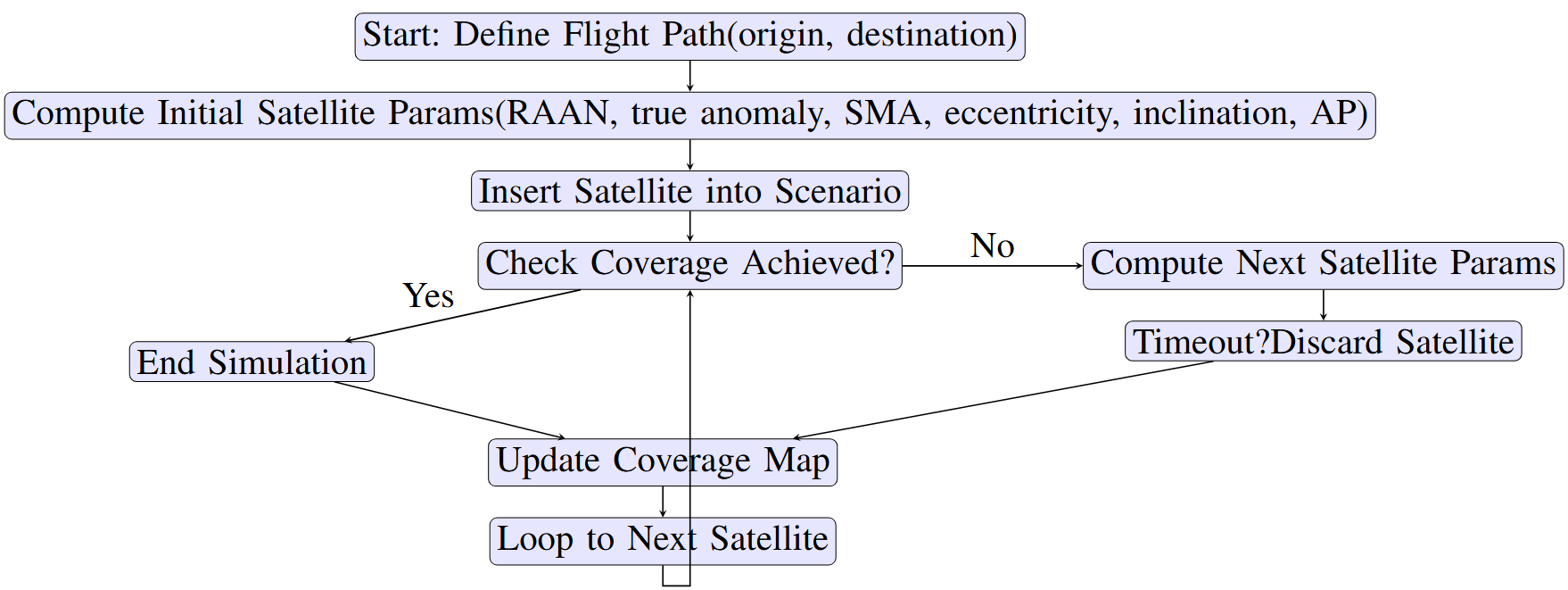}
\caption{Flowchart of the sequential satellite insertion process for LEO-based 5G coverage.}
\label{fig:satellite_flowchart}
\end{figure}

\vspace{-2mm}
%\subsection{LEO Satellite-Based 5G Connectivity}
\subsection{Sequential Satellite Insertion and Handover Mechanism}
\label{handover}

This approach mimics actual constellations like Starlink, using Right Ascension of the Ascending Node (RAAN) and true anomaly adjustments to cover the route sequentially, in which we introduce an adaptive series-based insertion mechanism. Satellites are dynamically added during simulation based on uncovered segments of the aircraft trajectory. The heuristic algorithm is shown in \reff{fig:satellite_flowchart}. The initial satellite position is determined from the flight origin and destination. After another function tracks coverage progress and triggers new insertions. A timeout filter discards satellites that fail to connect within 2 minutes, preventing simulation stalls. The framework is designed to be highly scalable, supporting the seamless integration of multiple satellite constellations and additional ground stations. This ensures robust connectivity even in densely trafficked air routes and adaptability to future advancements in satellite technology. The ability to dynamically add more satellites improves service reliability, reducing the dependency on any single network provider.

\subsection{In-Cabin 5G Distribution and Signal}
The in-cabin network uses phased-array antennas placed along the aircraft. A 3D Airbus A350 model and MATLAB ray-tracing engine simulate signal propagation and optimize antenna layout. The internal distribution system consists of small-cell networks that use beamforming to direct signals to specific user devices, reducing interference and improving data throughput. The network also utilizes multiple-input multiple-output (MIMO) technology to enhance spectral efficiency. By deploying intelligent beamforming and MIMO strategies, the system ensures that signal strength remains consistent throughout the aircraft, mitigating dead zones typically found in enclosed metal structures.

An internal network of distributed antennas within the aircraft ensures optimal 5G signal distribution. Transmitters use a 4$\times$8 Uniform Rectangular Array (URA) for beamforming; user equipment (UE) use a compact 2$\times$2 Uniform Linear Array (ULA). Ray-tracing techniques are employed to determine the most efficient placement of antennas, minimizing interference and ensuring maximum coverage throughout the cabin. Simulations apply the shooting and bouncing rays method with two reflections and 1$^{\circ}$ ray separation.

 %This distributed system ensures that each passenger receives high-quality service, regardless of their seating location.

\section{Proof of Concept} \label{sec:exp}

\begin{figure*}[!t]
\centering
\includegraphics[width=1\linewidth]{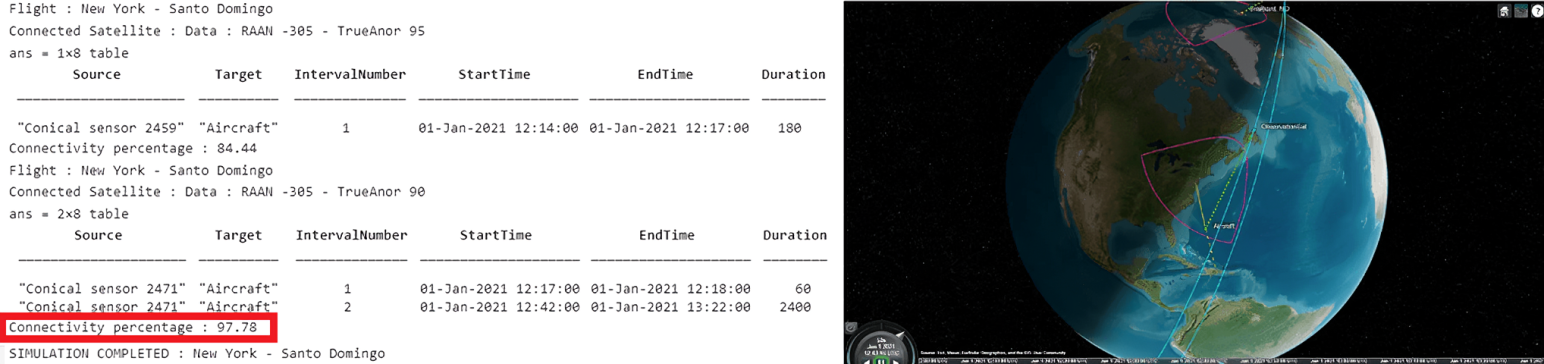}
\caption{Simulation environment: left part the connectivity percentage progress and in the right part the visual connection between satellite and the aircraft. Blue lines: satellite trajectories. Blue dots: satellites. Yellow: Aircraft trajectory. Green: Connection. Purple: Satellite view.}
\label{fig:globe}
\end{figure*}

A proof of concept (PoF) is carried out on a simulated flight from New York (NY) to Santo Domingo (SD). The measurements (M) considered in this PoC were: satellite handover efficiency, coverage reliability, and in-cabin 5G signal distribution. \reff{fig:globe} visually shows the simulation environment in two parts: (left) the results of one test simulating the flight by measuring the percentage of connectivity, and (right) while running the pre-defined flight trajectory (yellow dashed line) with the aircraft position (green point). As an example in \reff{fig:globe}, the test exhibits a successful teardown service below 2\% (red box in figure).  %The simulations were based on the previously developed Starlink satellite constellation model, successfully demonstrating improved connectivity along predefined flight paths.

\subsection{M1: Satellite Handover Efficiency}

One of the primary challenges in maintaining uninterrupted connectivity during flight is ensuring seamless handover between LEO satellites. The simulations implemented a dynamic handover algorithm for satellite positioning and aircraft trajectory. This algorithm continuously monitors real-time telemetry data from the aircraft and the satellite network. It dynamically adjusts beamforming parameters to ensure optimal alignment with the available satellite signal. The adaptive mechanism mitigates signal dropouts commonly occurring during satellite transitions and enhances the overall quality of service for passengers and onboard systems. The handover mechanism uses Simulink and MATLAB. The simulation iteratively adds satellites with adjusted orbital parameters—RAAN and true anomaly—to ensure seamless coverage. Inspired by the Starlink constellation, this method dynamically introduces new satellites as the aircraft transitions out of an existing coverage zone, optimizing signal continuity. Automated MATLAB functions track flight coverage and calculate initial satellite positions, enabling an adaptive handover process that minimizes signal loss and ensures near-seamless communication throughout the flight.

%The simulation results demonstrate that with the implementation of this handover mechanism, signal loss was reduced to below 0.5 (percentage) over the entire flight duration in some cases, ensuring a near-seamless experience. Additionally, the deployment of phased-array antennas on the aircraft played a crucial role in stabilizing the signal reception, allowing real-time adjustments to compensate for rapid changes in satellite position relative to the aircraft’s trajectory.

\subsection{M2: Impact of Additional LEO Satellites}

To enhance coverage reliability, simulations incorporated different densities of LEO satellites. The results clearly indicate that an increased number of satellites within a given flight corridor significantly boosts the probability of maintaining uninterrupted connectivity. The study explored various satellite deployment scenarios to evaluate their impact on service stability and quality, including minimal, standard, and dense configurations. We have evaluated two approaches: the parallel approach, where satellite values are inserted in the Simulink workflow, and the series approach, where a MATLAB file includes all satellite values. 

When deploying the series approach, our simulations confirmed that connectivity significantly improved along the aircraft flight. The results show that a parallel covering a transatlantic flight path resulted in a connectivity probability of only around 5\%, highlighting the limitations of a sparse satellite deployment. However, when a constellation of more satellites was introduced within the same corridor, the probability of maintaining a stable connection improved dramatically to up to 98\%. This finding underscores the importance of deploying a sufficiently dense LEO satellite network to support real-time handovers without noticeable service degradation. Additionally, integrating strategically positioned ground stations along high-traffic flight routes further reinforced signal reliability by providing alternative data pathways, particularly during satellite transitions or when atmospheric conditions affected satellite link stability.

Further refinements were made to test the insertion of satellites in series, ensuring each satellite aligned optimally within the constellation to maintain continuous coverage and a structure similar to the Starlink constellation. The study revealed that by strategically spacing the satellites along the flight path as the Starlink constellation, the aircraft maintained a more stable connection, reducing transition disruptions further. This approach proved especially beneficial in reducing the number of simultaneous satellite handovers. We have performed several tests in the North American area due to the current massive deployment of Starlink satellites. The configured parameter values were: RANN = $-170$, true anomaly: $390$, semi-major axis (SMA) = $7.2e6$ meters, eccentricity = $.05$ deg,
inclination = $70$ deg and argument of periapsis (AP) = $0$.     % degFigure \ref{fig:sim} shows the results of one test simulating the flight New York - Santo Domingo and measuring the percentage of connectivity. As observed, the coverage availability reached almost the 92\%. A total of four tests were performed. These tests were composed of two long flights: New York (NY)- Santo Domingo (SD) and New York- California (C) (round trip), hence, 4 different trips. The results of coverage availability was of: NY-SD - 91\%, SD-NY - 91\%, NY-C - 94.44\% and C-NY - 81\%. Indeed, we could reach a 89.36\% of coverage along the whole flight. These results are satisfactory considering that in average 10\% of the flight there are downtime experience \cite{wowfare_inflight_wifi}.   

%\begin{figure*}[!t]
%\centering
%\includegraphics[width=1\linewidth]{Figuras/handoverv2.png}
%\caption{Continuous connectivity along a flight New York - Santo Domingo with a successful teardown service below 9\%}
%\label{fig:sim}
%\end{figure*}

\subsection{M3: In-Cabin 5G Coverage Optimization}

\begin{figure}[!t]
\centering
\includegraphics[width=3.2in]{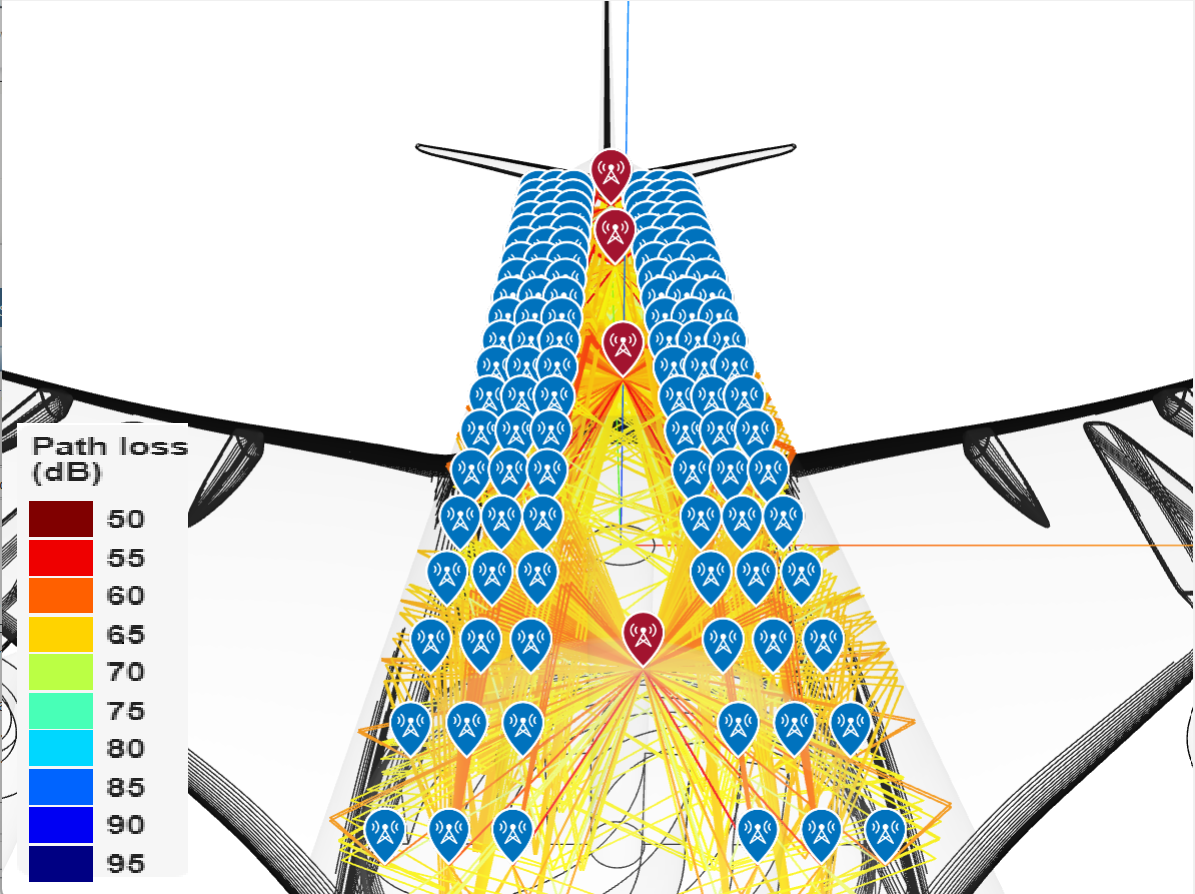}
\caption{In-Cabin distribution of four transmitters and 120 receivers (2 columns and 20 rows, 3 seats per column/row)}
\label{fig:raytracing}
\end{figure}

Ensuring high-quality connectivity inside the aircraft requires a sophisticated in-cabin distribution network. Ray-tracing simulations were conducted to analyze signal propagation, identify interference hotspots, and determine optimal antenna placement for maximum efficiency. The system design incorporated MIMO configurations and small-cell architectures to distribute the 5G signal evenly throughout the aircraft’s cabin, simulating the terminals of travelers located in their seats. To test the in-cabin 5G coverage, we used an aircraft 3D model and added transmitters and receiver nodes within it to simulate the raytracing behavior. \reff{fig:raytracing} displays how the transmitters (red nodes) are communicating to receivers (blue nodes) and the path loss for each trace. We have used four transmitters as a preliminary setup for further optimization of a number of nodes to cover all passengers' seats. We have used a typical employed large MIMO array 4×8 uniform rectangular array, to maximize beamforming and spatial multiplexing gains on the customer premise equipment (CPE). On the other side, we have chosen UEs with a simpler configuration, a 2×2 linear array, due to physical size and power constraints. This trade‐off enables the CPE to deliver high performance while keeping the UE compact and energy efficient \cite{3gpp38901}. We have used the standard frequency range 1 (FR1) of 5.8 GHz. Also, we defined a ray tracing propagation model using the shooting and bouncing rays method within a Cartesian coordinate system. We specify a low angular separation between rays, up to two reflections, and a metallic surface.   

\begin{figure}[!t]
\centering
\includegraphics[width=3.2in]{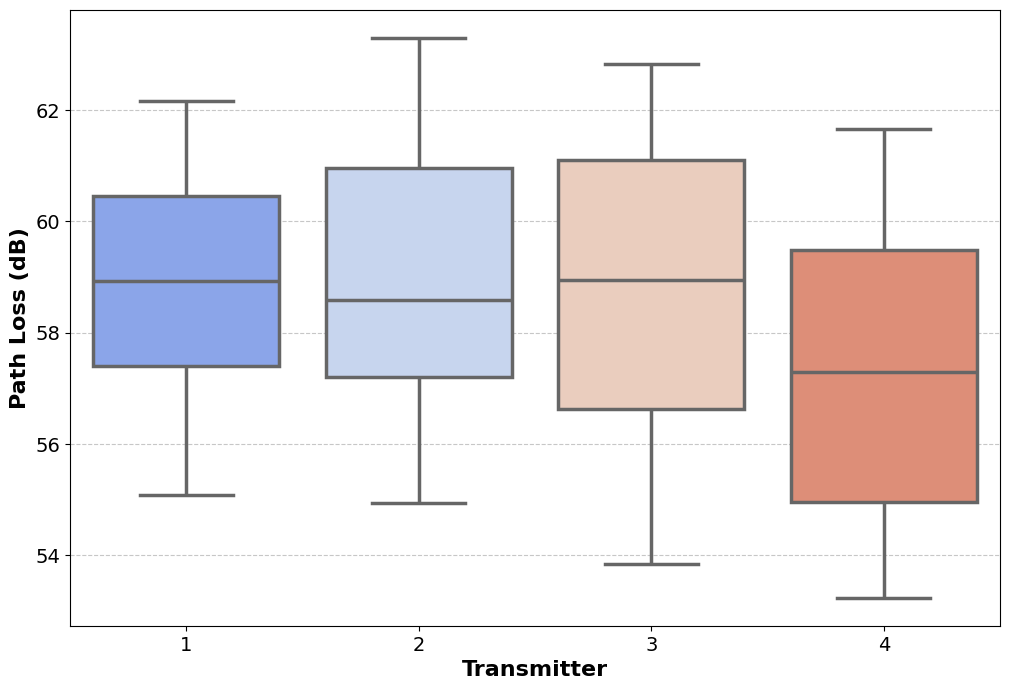}
\caption{Path loss vs Transmitter boxplot}
\label{fig:boxplot}
\end{figure}

%Previous simulations utilized ray-tracing models to assess in-cabin connectivity challenges, mainly focusing on signal attenuation due to the aircraft’s internal structure. 
The simulation outcomes reveal that deploying a network of strategically placed distributed antennas significantly reduced shadowing effects and signal attenuation. \reff{fig:boxplot} shows a boxplot with the path loss in dB per transmitter, where it can be observed how the mean values are similar among them, demonstrating a balanced quality along the plane.
By integrating advanced beamforming techniques, the system maintained a global path loss of 58.55 dB across all passenger seating areas. In the challenging environment of an airplane cabin transmitting 5G signals from LEO satellites (towards a CPE), this path loss value is primarily attributable to the combined effects of the aircraft's metallic fuselage attenuating signals, the transmitter-receiver distance, and the multipath propagation \cite{aerospace11070522}. Despite sophisticated beam-steering techniques, maintaining reliable signal quality through these multiple barriers while compensating for rapid relative movement between transmitter and receiver represents one of the most demanding scenarios in wireless communications, making even this elevated error rate a testament to the system's capabilities.  %These enhancements ensured a consistent connectivity experience for all passengers, regardless of their position within the cabin.

%%%%%%%%%%%%%%%%%%%%%%%%%%%%%%%%%%%
\section{Results Discussion}\label{sec:dis}
The results provide strong evidence supporting the viability of this approach in ensuring continuous 5G connectivity throughout commercial flight routes.
The simulations conducted in this study have yielded several important insights regarding satellite coverage for aircraft during flight and the quality of 5G connectivity inside the aircraft cabin. Regarding satellite coverage, the results demonstrate high availability, with a maximum coverage time of 98\% achieved for some flight routes.
This represents a substantial increase from the initial 5\% coverage, indicating that using multiple LEO satellites in a constellation configuration is promising for
providing near-continuous connectivity to aircraft. The automatic insertion of satellites based on the flight path proved effective in optimizing coverage, particularly for routes over areas with good Starlink infrastructure like the United States.
However, several challenges were encountered that prevented achieving 100\% coverage. The most significant issue was related to satellite synchronization and execution time in the
simulation environment. %Some satellite parameter combinations resulted in execution loops or timeouts, leading to potential coverage gaps. This highlights the need for further optimization of the simulation algorithms and possibly more computational resources to handle the complexity of a full satellite constellation.
The study also revealed the importance of carefully selecting orbital parameters like the RAAN and true anomaly to ensure optimal satellite
positioning relative to the aircraft's path. The series insertion method proved more flexible and efficient than parallel insertion, allowing for dynamic adjustment of the constellation as needed. Regarding in-aircraft coverage, the ray tracing simulations showed promising results in terms
of signal distribution. Using multiple transmitters strategically placed throughout the cabin helped achieve a more homogeneous coverage than a single transmitter setup. This approach mitigates issues related to signal attenuation over distance within the aircraft. The path loss analysis revealed generally good signal quality across most of the cabin, with average path loss value below 58 dB at least in one of the four transmitters. This indicates that transmitter placement is crucial in maintaining consistent signal quality. This suggests that further optimization of transmitter positions and possibly adding more transmitters could lead to even better coverage. 

\section{Conclusion} \label{sec:conc}

This study has demonstrated the feasibility and advantages of leveraging LEO satellite constellations to enable continuous 5G connectivity in aircraft. Our simulations achieved high coverage reliability through satellite deployment and adaptive handover mechanisms, with up to 97.78\% connectivity on transcontinental routes. Additionally, in-cabin 5G distribution was enhanced using ray-tracing techniques and transmitter placements across passenger areas. While challenges remain, particularly in fine-tuning satellite synchronization and improving in-cabin coverage, our findings establish a solid foundation for future advancements in satellite-based 5G networks for aviation. Future work will expand the study to additional constellations, optimize in-flight connectivity for seamless global coverage and combine both contributions providing additional output such as throughput and latency metrics.

% or


%\appendices

%\section{}

%\section*{Acknowledgment}


\ifCLASSOPTIONcaptionsoff
  \newpage
\fi



%\bibliographystyle{IEEEtran}
%\begin{thebibliography}{1}

%\end{thebibliography}
\bibliographystyle{IEEEtran}
\bibliography{bibtex/bib/main}
\vspace{-3em}
\begin{IEEEbiographynophoto}{Raúl Parada} received his B.Sc and M.Sc degree in Telecommunications Engineering from the Universitat Politècnica de Catalunya (UPC) in 2008, and from the Danmarks Tekniske Universitet (DTU) in 2012, respectively. And a Ph.D degree in Information and Communication Technologies %at the Departament de Tecnologies de la Informació i les Comunicacions (DTIC) 
from Universitat Pompeu Fabra (UPF) in 2016. Since 2015, he is Course Instructor with Universitat Oberta de Catalunya (UOC). He is a Senior Researcher %in the Sustainable Artificial Intelligence (SAI) research unit 
at the Centre Tecnològic de Telecomunicacions de Catalunya (CTTC).   \end{IEEEbiographynophoto}\vspace{-3.5em}
\begin{IEEEbiographynophoto}{\textbf{Victor Monzon Baeza}}
(SM'24) received the B.Sc., M.Sc., and Ph.D. (Hons.) degrees in electrical engineering from the University Carlos III of Madrid, Spain, in 2013, 2015, and 2019, respectively. Since 2019, he is Collaborator with Universitat Oberta de Catalunya. Currently, he is a Payload Architect Leader at Sateliot, Spain. \end{IEEEbiographynophoto}\vspace{-3.5em}
\begin{IEEEbiographynophoto}{\textbf{Carlos Horcajo Fernández de Gamboa}} received his B.S degree in sound and image engineering from the Universidad Politécnica de Madrid in 2018 and an M.Sc degree in telecommunications engineering from the Universitat Oberta de Catalunya in 2025. Currently, he is a technician engineer at Systelen, France.
\end{IEEEbiographynophoto}\vspace{-3.5em}
\begin{IEEEbiographynophoto}{\textbf{Rocío Serrano Camacho}} received her B.S degree in telecommunications technologies engineering from the Universidad de Alcalá in 2019 and an M.Sc degree in telecommunications engineering from the Universitat Oberta de Catalunya in 2024. Currently, she is a telecommunication engineer at Indra, Spain.
\end{IEEEbiographynophoto}\vspace{-3.5em}
\begin{IEEEbiographynophoto}{Carlos Monzo} (SM'25) received his BSc and MSc in Telecommunications Engineering, and his PhD degree about ``ICT and its management'', at Universitat Ramon Llull (URL) in 2001, 2003 and 2010, respectively. He is an associate professor and researcher at Universitat Oberta de Catalunya (UOC) since 2010. He has published several papers on the Internet of Things, signal processing, and e-learning.
\end{IEEEbiographynophoto}


%\end{IEEEbiography}


%\newpage



%\vfill


\end{document}